\documentclass[aip,amsmath,amssymb,
reprint,groupedaddress,onecolumn]{revtex4-2}

\usepackage{graphicx}
\usepackage{dcolumn}
\usepackage{bm}

\usepackage{mathtools}
\usepackage{lipsum}
\usepackage{hyperref}

\newcommand{\reffig}[1]{Fig.\,\ref{fig:#1}}

\newcommand{\refq}[1]{Eq.\,(\ref{eq:#1})}

\begin{document}

\title[Tapered Schroeder Complex]{Suppressing spectral edge effects in Schroeder's Harmonic Complex}
\author{Alessandro Alto\`e}
 \email{altoe@usc.edu}
\affiliation{Auditory Research Center\\ Caruso Department of
  Otolaryngology\\ University of Southern California\\ Los Angeles, CA
  90033
}%
  
\preprint{Author, JASA-EL}  
\begin{abstract} Schroeder's harmonic complexes are periodic, band-limited signals that are analogous to tones whose frequency increases or decreases over time. As such, they have been widely employed to study phenomena related to frequency-dispersion and frequency-modulation sensitivity in the auditory system. However, Schroeder's complexes also embed two steady ``frequency-fixed'' components. Because these components are easily audible, they may complicate interpretation of behavioral experiments. Here I present a variation of Schroeder's harmonic complex that largely suppresses these undesired steady components.
\end{abstract}
\maketitle

\section{Introduction}
In his seminal paper,\cite{schroeder1970} Manfred R. Schroeder described how to synthesize periodic, band-limited signals with approximately flat envelope (low crest factor). 
These signals commonly called Schroeder's harmonic complexes, can be thought of as tones whose frequency periodically increases (or decreases) from $f_1$ to $f_2$ over time.
As such, they have been broadly employed in hearing research, especially to study the effects of frequency-dispersion (e.g., frequency-dependent delay) in the auditory system. In particular, they have been broadly employed in behavioral studies\citep{kohlrausch1995,carlyon1997,summers1998,lentz2001,oxenhamdau2001,henry2023} and, to a lesser extent, in physiological studies.\cite{summers2003,recio2001schroeder,steenken2022,henry2023}

A largely overlooked issue with Schroeder's harmonic complexes, that makes them less than ideal to study frequency-dispersion, is that the signal components at the edge of their frequency band (e.g., at $f_1$ and $f_2$) do not have well-defined group delays and hence they are not well localized in time. As we will see below, Schroeder's harmonic complexes appear as the superposition of a tone whose frequency changes over time, with two fainter, steady tones.\cite{Note1} These steady signal components are audible for certain combinations of stimulus parameters, potentially playing an unrecognized (or undesirable) role for the interpretation of behavioral experiments that make use of these stimuli.\cite{Note2} 

Schroeder's complexes are closely related to frequency modulated (FM) sweeps, also widely used in auditory research,\cite{sinex1981,long2008,charaziak2026} that are tones whose instantaneous frequency increases (or decreases) from $f_1$ to $f_2$---these stimuli are obtained by modulating the frequency of a sinusoidal oscillator with e.g., a linear ramp. Indeed, Schroeder \cite{schroeder1970} synthesizes his harmonic complexes essentially by approximating periodic FM sweeps with a sum of sinusoids.
While FM sweeps that are made periodic do not manifest the problem noted with the Schroeder's harmonic complexes, the sudden change in their instantaneous frequency from $f_2$ to $f_1$ at the beginning of each period, produces a localized broadband response. In laymen's terms, periodic FM sweeps sound very much like the superposition of a tone whose frequency changes over time, with a click train.

Here, I propose a method to design periodic, band-limited stimuli that retain the time-varying frequency characteristic found in Schroeder's complexes and periodic FM sweeps, while suppressing undesirable steady and click-like signal components found in those stimuli.

{\bf Note}: it is not my intention to review the theory and history of Schroeder's harmonic complexes—the interested reader is referred to Schroeder's paper and related literature.\cite{schroeder1970,blachman1969,lentz2001}

\section{Schroeder's harmonic complexes}

Schroeder's harmonic complexes are band-limited ($f \in [f_1,f_2]$) periodic signals with repetition rate $r$ and period $T=1/r$. They are the linear superposition of $N$ harmonics
\begin{equation}
s(t)=\sum_{n=0}^{N}A_n\sin(2\pi r(j+n)t+\phi_n), \label{eq:sch1}
\end{equation} 
with $f_1/r=j \in \mathbb{N}$, $f_2/r=l \in \mathbb{Z}$, and $N=l-j+1$. The phase terms $\phi_n$ are selected so that each harmonic has a specific group delay; the coefficients $A_n$ control the harmonics' amplitude.

Schroeder designed the phase term specifically to minimize the waveform's crest factor. He achieved that goal essentially by spacing the group delays of the different harmonics across the stimulus period, in a way that depends on the harmonics' relative power \mbox{$p_k=A_k^2/\sum_n A_n^2$}. The resulting equation for calculating the phase is
\begin{equation}
	\phi_n=\phi_0-2\pi\sum_{k=0}^{n-1}\sum_{i=0}^{k}p_i. \label{eq:sch2}
\end{equation}
Note that in the literature the name ``Schroeder complex'', is typically used to describe variations of this stimulus in the special case that all harmonics have the same amplitude (e.g., $A(f)=1$ with $f\in[f_1,f_2]$).
In this case, assigning $\phi_0=0$ yields
\begin{equation}
\phi_n=-\pi \frac{n(n+1)}{N}, \label{eq:linear_sch}
\end{equation}
i.e., the stimulus has a quadratic phase curvature and constant amplitude in $[f_1,f_2]$. This means the resulting signal has approximately flat envelope and instantaneous frequency increasing linearly over time from $f_1$ to $f_2$.\cite{schroeder1970} Popular variations of these stimuli include additional parameters to controls the signal's crest factor and phase curvature.\cite{lentz2001}

The top section of \reffig{1} shows an example Schroeder's harmonic complex, where $f_1=1.6$ kHz, $f_2=6.4$ kHz and $r=50$ Hz.\cite{Note3} From left-to-right, the figure shows the signal's waveform, its amplitude spectrum, and its time-frequency representation through a spectrogram. The spectrogram of the Schroeder's complex is very interesting. It is dominated by diagonal ``sawtooth-like'' lines, that represent a signal component whose frequency periodically increases linearly over time. However, the spectrogram also shows two not-so-faint horizontal lines centered around $f_1$ and $f_2$, meaning that the signal comprises two steady ``frequency-fixed'' components of frequencies $f_1$ and $f_2$. As anticipated in the introduction, when listening to Schroeder's harmonic complexes, especially at low modulation rates, these two steady components stand out and are audible.\cite{Note4}

The presence of these frequency-fixed components can be understood by noting that the group delay $\tau$ (within one signal period) can be calculated from the signal's phase curvature for all harmonics with frequency $f \in (f_1,f_2)$:
\begin{equation}
\tau_n=-\frac{1}{2\pi}\frac{d\phi}{df}=-\frac{1}{2\pi r}\frac{d\phi_n}{dn}= T\Big(\frac{n}{N}+\frac{1}{2N}\Big) \qquad \text{for } 0<n<N, \label{eq:gd_sch}
\end{equation}
where the reader must forgive the abuse of notation.\cite{Note5}
However, because the signal phase is undefined for $f<f_1$ and for $f>f_2$, the phase curvature and hence the group delay cannot be calculated at $n=0$ and $n=N$.

\section{Periodic FM Sweeps}
A frequency-modulated (FM) sweep is obtained by modulating the frequency of a sinusoidal oscillator such that its instantaneous frequency ``sweeps'' from $f_1$ to $f_2$ over the signal's period $T$.
The sinusoidal oscillator can be defined as
\begin{equation}
s(t)=\sin(\phi(t)), \label{eq:chirp}
\end{equation}
where $\phi(t)$ indicates the instantaneous phase of the oscillator. The oscillator's instantaneous frequency is $f_i=\frac{1}{2\pi}\frac{d\phi}{dt}$.  In case of a sweep where the instantaneous frequency increases linearly over time from $f_1$ to $f_2$ over the period $T$:
\begin{equation}
	f_i(t)=\frac{1}{2\pi}\frac{d\phi}{dt}=f_1+\frac{f_2-f_1}{T}t \qquad \text{for } 0\le t<T,
\end{equation}
and the instantaneous phase is
\begin{equation}
\phi(t)=2\pi \int f_i dt=2\pi \Big(f_1t+\frac{f_2-f_1}{2T}t^2\Big) \qquad 0\le t \le T, \label{eq:phase}
\end{equation}
where, for simplicity, we have assumed that $\phi(0)=0$.

These sweeps are made periodic simply by ``cycling'' $\phi$ over one period:
\begin{equation}
\phi(t)=2\pi\Big( f_1t+\frac{f_2-f_1}{2T}(t-T\lfloor\frac{t}{T}\rfloor)t\Big). \label{eq:phase_periodic}
\end{equation}
It is important to appropriately select the frequency limits and period so that $\phi(T)=2\pi k$ with $k\in \mathbb{N}$, to avoid waveform's discontinuities at the beginning of each cycle.
 
The middle panel of \reffig{1} shows an example of a linear FM sweep. This signal shares many similarities with the Schroeder's harmonic complex in the panel above, but also shows some key differences. While the Schroeder's complex has a perfectly flat frequency spectrum within $f_1$ and $f_2$, but presents visible ripple in its envelope, the opposite is true for the FM sweep: its envelope is perfectly flat, but its amplitude spectrum shows visible ripples. 

More interesting, the horizontal lines visible in the spectrogram of the Schroeder's complex, are absent from the spectrogram of the FM sweep. However, the latter shows vertical lines at the beginning of each cycle, indicating the presence of a highly localized broad-band signal component. 
The presence of such component is explained by that, even though we ensured that the signal waveform and phase are continuous at the beginning of each cycle, the FM sweep instantaneous frequency is not, suddenly jumping from $f_2$ to $f_1$. Despite all the measures one can take to ensure continuities at the beginning of each cycle, the resulting periodic FM sweep will always sound rather ``clicky''.

\section{Tapered Schroeder complex}
I now propose a stimulus design method, based on Eqs.\,(\ref{eq:sch1},\ref{eq:sch2}), that largely suppresses stationary and broadband components present in Schroeder's complexes and periodic FM sweeps respectively. The idea is to slightly extend the Schroeder's harmonic complex band, to avoid discontinuities at its edges. In particular, I let the harmonics' amplitude to be constant within the signal band $[f_1,f_2]$. Outside this band, I impose the amplitude to decay with a simple tapering function.\cite{Note6} Given the signal band $[f_1,f_2]$ and rate $r$,  the harmonics amplitudes are calculated as
\begin{equation}
	\begin{cases}  \label{eq:tapered}
		A(f)=1 \quad &  f_1<f< f_2, \\
		A(f)=(f/f_1)^M \quad &{f\le f_1},\\ 
		A(f)=(f/f_2)^{-M} \quad &f\ge f_2. 
	\end{cases}
\end{equation}
where $f=nr$ indicates the frequency of the $n$-th harmonic, and $M$ indicates the order of the tapering function---the amplitude decreases at a rate of $6M$ dB/octave outside the signal's band.\cite{Note7} The phase of each harmonic is then determined by \refq{sch2}. 

I call the resulting stimulus a ``tapered Schroeder complex''. Its realization using a 16-th order tapering function [$M=16$ in \refq{tapered}] is shown in the bottom panel of \reffig{1}. Note that in this realization, $f_1$ and $f_2$ have been slightly adjusted so that the resulting signal retains the same equivalent rectangular bandwidth of the Schroeder's harmonic shown in the top panel. The tapered Schroeder complex waveform and amplitude spectrum are very similar to those of the Schroeder's harmonic complex, but with less visible ripples in the envelope, and less sharp spectral edges.

More importantly, the spectrogram of the tapered Schroeder complex does not show the vertical and horizontal lines present in those of the Schroeder harmonic complex and FM sweep, respectively. A little vertical and horizontal ``smearing'' are present at the beginning and end of each stimulus' cycle. The parameter $M$, by controlling the sharpness of the spectral edges, in practice controls the trade-off between vertical and horizontal smearing.\cite{Note8}

According to what is shown in the spectrogram, when listening to tapered Schroeder complexes there are no prominently audible steady “frequency-fixed” components as there are when listening to Schroeder complexes, nor is there a prominent click-like sound at the beginning of each cycle, as stands out when listening to FM sweeps. This is true even when listening to signals with low modulation rates, where these components are more easily heard. Examples audio files are provided in the software repository.

\section{Discussion and Conclusion}
I presented the design of a band-limited signal, named ``tapered Schroeder complex'', with low crest-factor and with well defined group delays across the entire signal band. The proposed signal is a modification of the Schroeder's harmonic complex,\cite{schroeder1970}  that avoids the sharp spectral edges present in the original stimulus. It allows one to synthesize a signal that is well described by a tone whose frequency periodically increases over time, while suppressing steady and click-like components that are present in Schroeder's harmonic complexes and periodic FM sweeps, respectively.

\begin{figure}
\includegraphics[width=\textwidth]{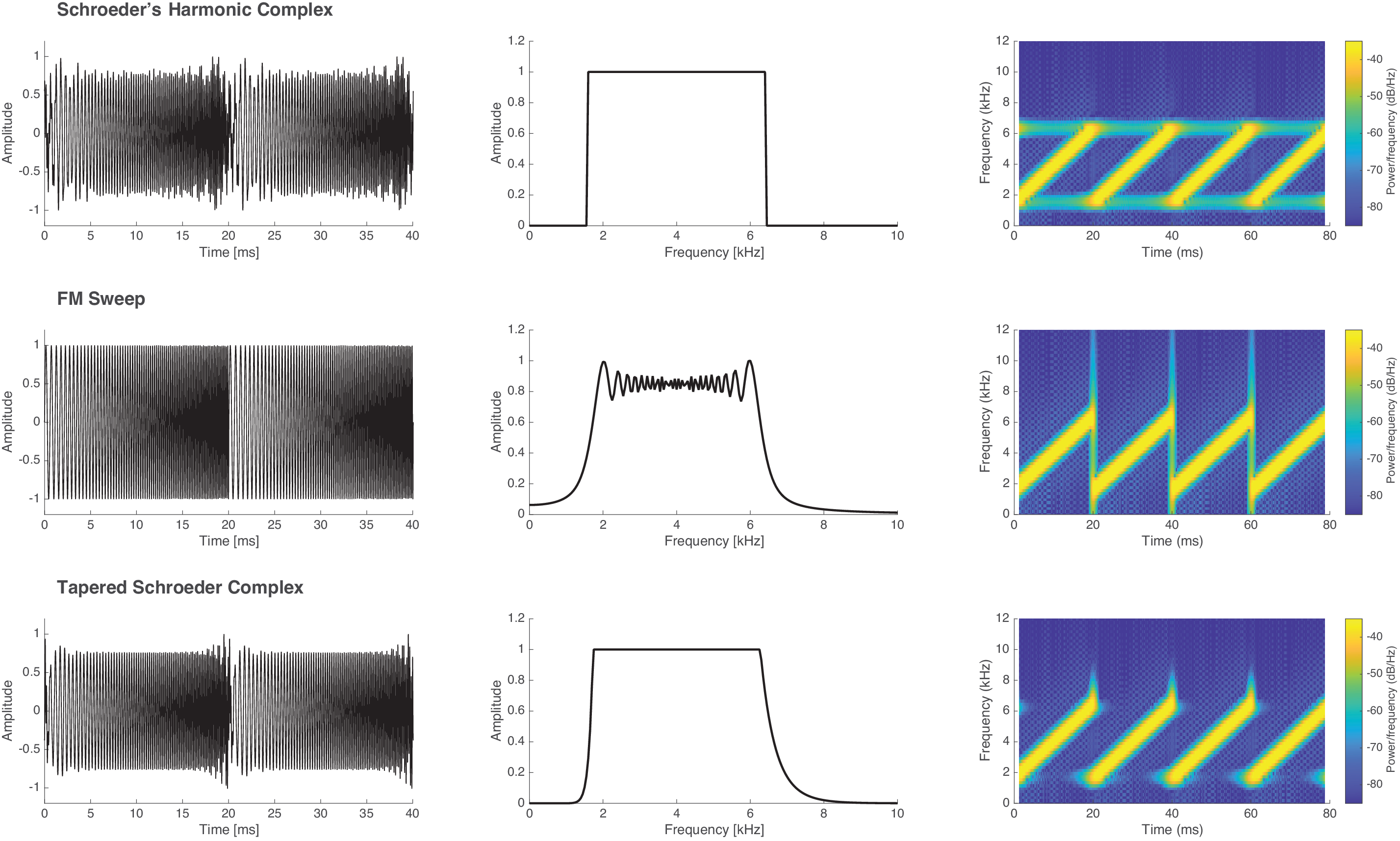}
\caption{Waveform (left), Amplitude spectrum (center) and spectrogram (rigth) of different periodic signals with low crest factor. Top: Schroeder's harmonic complex. Middle: periodic frequency-modulated (FM) sweep. Bottom: Tapered Schroeder complex (the stimulus proposed here). }\label{fig:1}	
\end{figure}

Similar results could likely have been obtained by modifying the periodic FM sweep in order to ``fade in'' and ``fade out'' the initial and final portion of each cycle, with the goal of suppressing the unwanted broadband response (middle panel of \reffig{1}). I decided to not follow that route for different reasons, the main one being that the Schroeder's algorithm gives the designer full control of the group delays at each frequency component. Conversely, the FM sweep gives the designer full control of the instantaneous frequency's trajectory, but not of the group delay. While group delay and instantaneous frequency's trajectory are closely related, they are not the same; the error in mapping one into the other increases with an increasing modulation rate.\cite{blachman1962} When studying frequency-dispersion, the group delay, not the instantaneous frequency, is the important stimulus parameter.

Note that with only minor modifications the method presented here is suitable to design stimuli with a variety of frequency trajectories. For example, it can be used to design a signal whose frequency increase (or decreases) exponentially over time (akin to the periodic version of the ``log-sweep'' \cite{farina2000}), or to improve a popular variation of the Schroeder's complex---known as ``Schroeder-phase harmonic''---that includes an additional parameter ($C$) to control crest factor and frequency trajectory.\cite{lentz2001} Scripts to implement these stimuli are provided in the software repository.

\section*{Acknowledgment}
I thank Karolina K. Charaziak and Christopher A. Shera for their support, and for their comments on the manuscript. Supported by NIH/NIDCD Grants R01 DC003687 (CAS) and R01 DC022942 (KKC). 
\section*{Author Declaration}
\subsection*{Conflict of interest} 
The author has no conflicts to disclose.
\section*{Data Availability}
The code used for the stimulus design is available at \url{https://github.com/AuditoryPhysicsGroup}.

\end{document}